\documentclass[twocolumn]{aastex63}
\usepackage{amssymb}
\usepackage{array}
\usepackage{multirow}
\usepackage{bold-extra}
\AtBeginDocument{
}

\newcommand{\n}{\nodata}

\received{2020 January 3}
\revised{2020 March 16}
\accepted{2020 April 2}

\submitjournal{ApJ}

\shorttitle{The origin of high-energy neutrinos in AGN}
\shortauthors{Plavin et al.}
\graphicspath{{./}{figures/}}

\begin{document}

\title{Observational evidence for the origin of high-energy neutrinos in parsec-scale nuclei of radio-bright active galaxies}

\correspondingauthor{Alexander Plavin}
\email{alexander@plav.in}

\author{Alexander Plavin}
\affiliation{Astro Space Center of Lebedev Physical Institute, Profsoyuznaya 84/32, 117997 Moscow, Russia}
\affiliation{Moscow Institute of Physics and Technology, Institutsky per. 9, Dolgoprudny 141700, Russia}

\author{Yuri Y. Kovalev}
\affiliation{Astro Space Center of Lebedev Physical Institute, Profsoyuznaya 84/32, 117997 Moscow, Russia}
\affiliation{Moscow Institute of Physics and Technology, Institutsky per. 9, Dolgoprudny 141700, Russia}
\affiliation{Max-Planck-Institut f\"ur Radioastronomie, Auf dem H\"ugel 69, 53121 Bonn, Germany}

\author{Yuri A.~Kovalev}
\affiliation{Astro Space Center of Lebedev Physical Institute, Profsoyuznaya 84/32, 117997 Moscow, Russia}

\author{Sergey Troitsky}
\affiliation{Institute for Nuclear Research of the Russian Academy of Sciences, 60th October Anniversary Prospect 7a, Moscow 117312, Russia
}

\begin{abstract}
Observational information on high-energy astrophysical neutrinos is being continuously collected by the IceCube observatory. However, the sources of neutrinos are still unknown.
In this study, we use radio very-long-baseline interferometry (VLBI) data for a complete VLBI-flux-density limited sample of active galactic nuclei (AGN). We address the problem of the origin of astrophysical neutrinos with energies above 200~TeV in a statistical manner. It is found that AGN positionally associated with IceCube events have typically stronger parsec-scale cores than the rest of the sample. The post-trial probability of a chance coincidence is 0.2\,\%. We select the four strongest AGN as highly probable associations: 3C\,279, NRAO~530, PKS~1741$-$038, and PKS~2145$+$067. Moreover, we find an increase of radio emission at frequencies above 10~GHz around neutrino arrival times for several other VLBI-selected AGN on the basis of RATAN-600 monitoring. The most pronounced example of such behavior is PKS~1502$+$106.
We conclude that AGN with bright Doppler-boosted jets constitute an important population of neutrino sources. High-energy neutrinos are produced in their central parsec-scale regions, probably in proton-photon interactions at or around the accretion disk. Radio-bright AGN that are likely associated with neutrinos have very diverse $\gamma$-ray properties suggesting that $\gamma$-rays and neutrinos may be produced in different regions of AGN and not directly related. A small viewing angle of the jet-disk axis is, however, required to detect either of them.
\end{abstract}


\section{Introduction} \label{s:intro}

Extraterrestrial neutrinos with energies $E\gtrsim 50$~TeV have been convincingly observed by the IceCube experiment since 2012 (\citealt{IceCubeFirstPeV,IceCubeFirst26}; for the most recent updates see \citealt{IceCubeICRC2019}). In 2019, the 
Baikal--GVD (Gigaton Volume Detector) experiment reported on the first few $E>100$~TeV neutrino candidates \citep{Baikal2019ICRC}, opening the way to test the IceCube observation from the Northern hemisphere. Indications to the astrophysical high-energy neutrino flux were also found by the 
ANTARES experiment \citep{ANTARES2019ICRC}. Despite these various observations, the origin of the energetic astrophysical neutrinos remains unknown (for a review, see, e.g., \citealt{AhlersHalzen}). Since the arrival directions of the neutrinos do not demonstrate any significant Galactic anisotropy \citep[see, e.g.,][]{ST-Gal, IceCubeANTARES-Gal}, their origin in extragalactic sources is often assumed. Active galactic nuclei (AGN) were discussed as potential neutrino emitters long before the neutrino detection (\citealt{BerezinskyNeutrino77}; see also, e.g., \citealt{Eichler1979,BerezGinzb} for subsequent early studies). Further interest in this class of sources was sparked by the observation of a $\gamma$-ray flare of the blazar TXS~0506+056 in a directional and, to a certain precision, temporal coincidence with the neutrino event 170922A detected by IceCube \citep{IceCubeTXSgamma}. This event was supplemented by an excess of lower-energy neutrinos from the same direction found in the archival data \citep{IceCubeTXSold}. Nevertheless, the origin of the entire population of the observed neutrinos in AGN is strongly constrained \citep[see, e.g.,][and \autoref{s:discussion:previous} of the present paper]{MurWaxMultiplets,Murase10percent,MuraseCombined}. 

However, joint analyses of the IceCube data sets obtained with various experimental techniques reveal a possibility that the observed astrophysical neutrino flux is formed by two distinct components, a softer one dominating the flux at $E\sim 50-100$~TeV and a harder one, which is important above $E\sim 200$~TeV \citep{Vissani2comp,AhlersHalzen}. Strong constraints on the origin of the entire population of neutrinos in active galaxies 
are relaxed for the hard component considered alone, so the origin of the dominant part of the observed neutrinos above $\sim$200~TeV in powerful AGN probably remains the best option. In the present study, we concentrate on this higher-energy component of the neutrino flux.

It is usually assumed that high-energy neutrinos are produced in decays of charged $\pi$ mesons, which are born as secondary particles in interactions of energetic protons with ambient matter or radiation. The acceleration of such protons and the presence of sufficiently abundant targets are, therefore, the key conditions for the neutrino production. In principle, they may be realized in various parts of AGN, and two general classes of models are considered with the neutrino production zone located either in the central (accretion disk, jet launching and acceleration region, broad-line region) or in the extended (kiloparsec-scale jets, blobs, lobes, hot spots) parts of a galaxy; see, e.g., reviews by \citet{Murase-rev,Meszaros-rev,Boettcher-rev,Cerruti-rev} and the references therein. It is a non-trivial task to distinguish between these two scenarios observationally because the poor angular resolution of astronomical instruments, especially those working at high energies, prevents one from direct localization of the regions where the radiation co-produced with neutrinos comes from. In addition, a low directional resolution of neutrino experiments and a high rate of atmospheric background events make the association of detected neutrinos with particular candidate sources challenging.

The aim of the present work is to alleviate these difficulties and to obtain direct observational evidence in favor of one of the scenarios. To distinguish between central and outer parts of active galaxies, we use very-long-baseline interferometric (VLBI) radio observations capable to resolve central parsecs of AGN even at cosmological distances, whereas the problem of source associations is addressed by a statistical approach. 
Note that the accretion disk is invisible in the radio, with the jet acceleration and collimation zone being resolved only for the nearby AGN \citep{2019arXiv190701485K}. However, activities observed in the apparent jet base by VLBI with a typical resolution in the plane of the sky of about 1~pc are shown to be good tracers of what is happening in and around the nucleus \citep[e.g.,][]{2002Natur.417..625M,2010ApJ...722L...7P}.

The rest of the paper is organized as follows. In \autoref{s:data}, we introduce the data sets used in our analysis: the IceCube neutrino events (\autoref{s:data_icecube}), the VLBI observations (\autoref{s:data_vlbi}), and the radio monitoring archive (\autoref{s:data_radio}). Section~\ref{s:analysis} presents the description and the results of the performed statistical analyses. In \autoref{s:discussion}, we compare our results with the previous studies and briefly discuss their implications for models of high-energy astrophysical neutrino production. We summarize our conclusions in \autoref{s:summary}.

\section{Data} \label{s:data}
\subsection{IceCube Events} \label{s:data_icecube}

\begin{table*}
\caption{IceCube high-energy neutrino events used in our analysis
\label{t:icecube_events}
}
\begin{center}
\begin{tabular}{cccrrrrrrc}
\hline\hline
Date & Category & $E$ & RA & \multicolumn{2}{c}{RA Error} & DEC & \multicolumn{2}{c}{DEC Error} & Reference \\
 & & (TeV) & ($\degr$) & \multicolumn{2}{c}{($\degr$)} & ($\degr$) & \multicolumn{2}{c}{($\degr$)} & \\
(1) & (2) & (3) & (4) & (5) & (6) & (7) & (8) & (9) & (10) \\
\hline
2009-08-13  & MUONT       & 480         & $29.51$     & $+0.40$     & $-0.38$     & $1.23$      & $+0.18$     & $-0.22$     & \citealt{1607.08006} \\
2009-11-06  & MUONT       & 250         & $298.21$    & $+0.53$     & $-0.57$     & $11.74$     & $+0.32$     & $-0.38$     & \citealt{1607.08006} \\
2010-06-23  & MUONT       & 260         & $141.25$    & $+0.46$     & $-0.45$     & $47.80$     & $+0.56$     & $-0.48$     & \citealt{1607.08006} \\
2010-09-25  & MUONT       & 460         & $266.29$    & $+0.58$     & $-0.62$     & $13.40$     & $+0.52$     & $-0.45$     & \citealt{1607.08006} \\
2010-10-09  & EHEA        & 660         & $331.09$    & $+0.56$     & $-0.72$     & $11.10$     & $+0.48$     & $-0.58$     & \citealt{https://doi.org/10.21234/b4ks6s,1607.08006} \\
\hline
\end{tabular}
\end{center}
\tablecomments{
The set of all 56 IceCube events selected according to our criteria, see \autoref{s:data_icecube} for details and category notations.
This is a sample of five rows only, the complete table is available electronically.}
\end{table*}

IceCube detects high-energy neutrino events of two types: cascades and tracks. The former are seen as showers that develop within the detector volume; the energy of the primary neutrino is determined relatively well but the arrival direction is uncertain. For the latter, the situation is the opposite: relatively narrow tracks pass through the detector; hence the angular resolution is normally of the order of $1\degr$, but some part of the energy of secondary particles is left outside the instrumental volume and the primary particle energy is determined with large uncertainties.
In the present study, we concentrate on the track events because of their better angular resolution.
We are interested in neutrinos with estimated energies $E\gtrsim 200$~TeV because it is the value above which, assuming two flux components, the hard-spectrum component starts to dominate. This can be seen, for instance, by comparison of the best-fit spectra obtained by IceCube from the analysis of starting events (more sensitive at lower energies) and of Northern-hemisphere muon tracks (more sensitive at higher energies), as reported by \citet{IceCubeICRC2019}. Remarkably, this value $E=200$~TeV is also the threshold value for some published IceCube Northern-hemisphere muon track data sets \citep[p.~30]{IceCube-1607.08006, IceCube-1710.01191}, which provides an additional technical motivation for this cut. Therefore, we fix the condition $E\ge 200$~TeV for all the tests discussed below. A study of validity of our conclusions for less energetic neutrinos is beyond the scope of the present paper. 

The largest published IceCube data set of high-energy track events is given by Extremely High Energy (EHE) alerts and alert-like (EHEA) events. This data set includes events that passed the selection criteria \citep{IceCubeOldAlerts} for the EHE type alerts issued by IceCube between July~2016 and May~2019. The list of events before September 2017, including early events that arrived before the launch of the alert system but satisfied the same criteria, is published online\footnote{\url{https://icecube.wisc.edu/science/data/TXS0506_alerts}} \citep{IceCubeEHEAcatalog}.
The details of similar events observed after September~2017 are available through the Gamma-ray Coordinates Network\footnote{\url{https://gcn.gsfc.nasa.gov/gcn_main.html}} (GCN) and Astrophysical Multimessenger Observatory Network\footnote{\url{https://www.amon.psu.edu}} (AMON) notices\footnote{\url{https://gcn.gsfc.nasa.gov/amon.html}}, see also IceCube Catalogue of Astrophysical Neutrino Candidates\footnote{\url{https://neutrino-catalog.icecube.aq}}.  For one event, we use the detailed information from \citet{IceCubeTXSgamma}.  By construction, the EHEA events have a good angular resolution (the 90\% containment area on the celestial sphere $\Omega_{90}<10$~sq.~deg) and high estimated energies (certainly above 200~TeV). There are 33 events in this EHEA sample. 

In order to use the largest available sample of highest-energy neutrino events of similar quality, we supplement the EHEA sample with 23 more events satisfying the following criteria: 
(i) track morphology,
(ii) $E>200$~TeV,
(iii) $\Omega_{90} < 10$~sq.~deg.
These events were selected from all the other publicly available IceCube event lists. They include High Energy Starting Event (HESE) alerts and alert-like events (HESEA), ``GOLD'' and ``BRONZE'' alerts from \citet{IceCubeEHEAcatalog} and GCN/AMON, HESE lists from \citet[p.~54]{IceCube-1405.5303,IceCube-1510.05223,IceCube-1710.01191}, and Northern-hemisphere muon track (MUONT) event lists from \citet[p.~30]{IceCube-1607.08006,IceCube-1710.01191}. For a few HESE alerts, the estimated energy of the neutrino has not been published; we then used the deposited charge (number of photoelectrons) divided by 100 as a proxy for the energy in TeV, cf.~\citet{IceCube-1405.5303}. Following \citet{corr-1601.06550Resconi}, we drop one MUONT event that was retracted. Note that some MUONT events appear in the EHEA list as well; we use the information from a more recent EHEA catalog for them.

For the IceCube events, coordinate-wise intervals with 90\% statistical coverage are reported in the published data we use. In addition, there exist unpublished systematic errors in the determination of the arrival directions, related in particular, but not exclusively, to the lack of knowledge of ice properties. These errors depend not only on the arrival direction but also on the part of the installation where the neutrinos land and are, therefore, hard to model. With the exception of a few events --- see, e.g.,\ \citet{Kankare:2019bzi} --- only statistical errors are provided for the published IceCube arrival directions; whereas for good-resolution events the contribution of these systematic errors can be important. The absolute IceCube pointing error was estimated by \citet{MoonShadow} as $\lesssim 0.2\degr$; however, the same paper states explicitly that smaller or larger errors may correspond to the events selected in particular neutrino analyses. Further, a contribution to the systematic error comes from the choice of the reconstruction procedure and may be estimated by comparison between the arrival directions of one and the same event obtained with different analyses. We found seven events whose arrival directions were published both in the EHEA and MUONT analyses, see the references above; the mean difference between the arrival directions in these two reconstruction was $\approx 0.25\degr$. Having no systematic errors published, we use as a guidance the published  IceCube upper limit of $1.0\degr$ \citep{IceCubeFirst26} on the systematic uncertainty of the arrival directions of high-energy muon tracks and further refine this value by means of the procedure defined in \autoref{s:analysis}.

\begin{figure*}[t]
\includegraphics[width=1.0\textwidth]{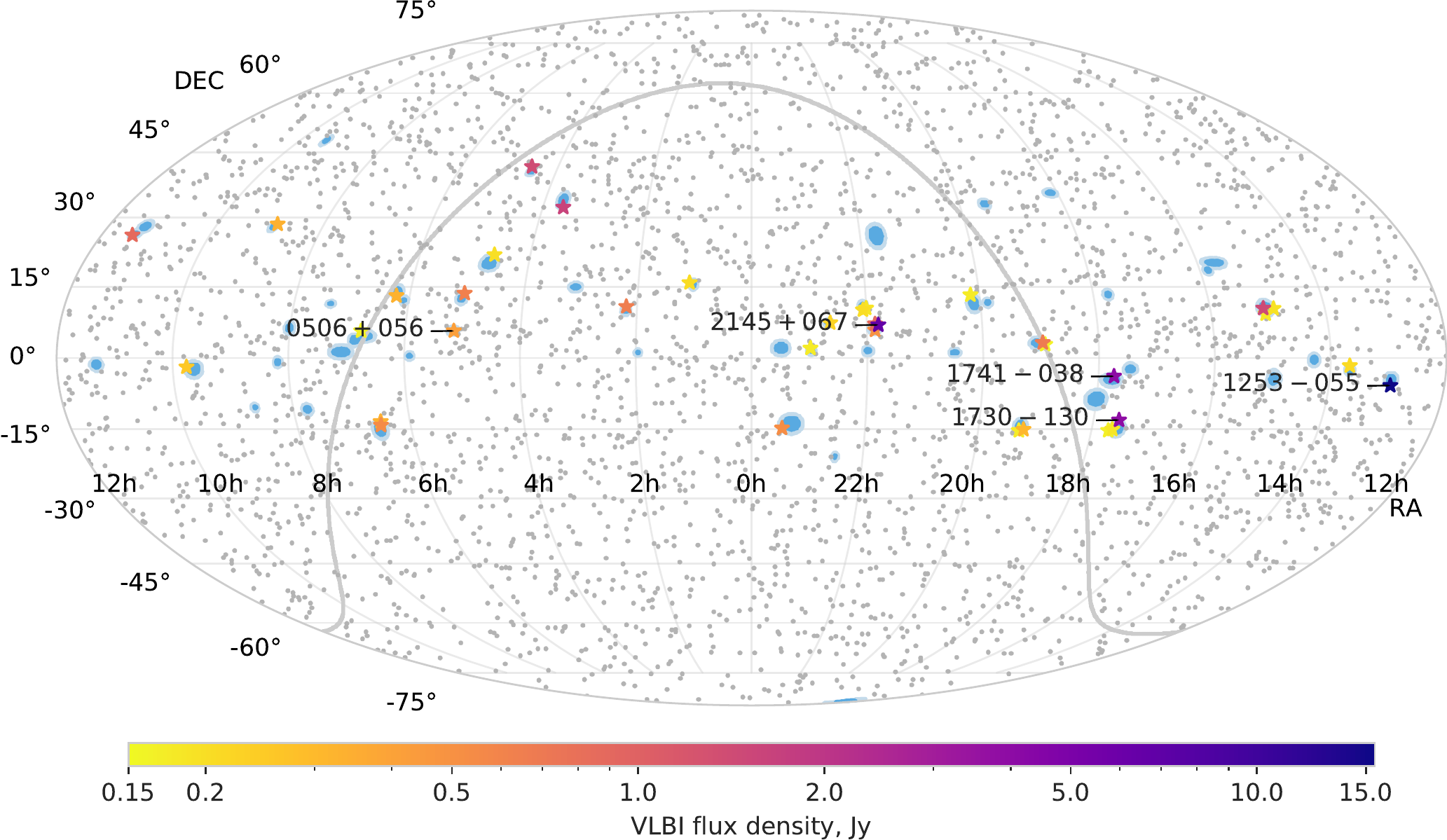}
\caption{
\label{f:skymap}
IceCube event locations on the sky, represented by blue ellipses. Dark blue ellipses are the original reported positional error regions, light blue ones are enlarged to account for unknown systematics according to our analysis, see \autoref{s:an_avgflux} for details and \autoref{s:data_icecube} for the event sample selection. Stars represent all AGN within neutrino error regions from our complete VLBI sample of AGN. Color represents the 8~GHz flux density integrated over the VLBI images of these AGN. Members of the complete 8-GHz VLBI sample down to 150~mJy located outside the ellipses are shown by grey dots.
The shown object names denote four AGN with the strongest parsec-scales jets that are the most probable neutrino associations according to our analysis: \object{1253$-$055} (\object{3C\,279}), \object{1730$-$130} (\object{NRAO~530}), \object{PKS~1741$-$038}, and \object{PKS~2145+067}. We also show the location of the first neutrino association \object{TXS~0506+056}.}
\end{figure*}

Therefore, our sample of the IceCube high-energy neutrinos includes 56 events with $E>200$~TeV, known arrival directions, 90\% confidence level (CL) statistical uncertainty ellipses on the celestial sphere, and arrival times. These events are listed in \autoref{t:icecube_events} and shown in \autoref{f:skymap}. Note that a significant part of the events is not astrophysical: even at high energies, the atmospheric background is essential. For instance, the expected fraction of non-astrophysical events in the EHEA sample, assuming $E^{-2}$ astrophysical spectrum, is $32$\% \citep{IceCubeOldAlerts}; for a softer assumed spectrum or for other event classes the background contribution is even larger. We also note that up till now neither Baikal-GVD nor ANTARES have published detailed information on track events above 200~TeV.

\subsection{VLBI Observations of AGN}
\label{s:data_vlbi}

For our analysis, we used 8~GHz VLBI observations compiled in the Astrogeo\footnote{\url{ http://astrogeo.org/vlbi_images/}} database, comprising the visibility data and images acquired from geodetic VLBI observations \citep{2009JGeod..83..859P,2012A&A...544A..34P,2012ApJ...758...84P}, the Very Long Baseline Array (VLBA) calibrator surveys (VCS; \citealt{2002ApJS..141...13B,2003AJ....126.2562F,2005AJ....129.1163P,2006AJ....131.1872P,2007AJ....133.1236K,2008AJ....136..580P,VCS9,2016AJ....151..154G}), together with other 8~GHz global VLBI, VLBA, EVN (the European VLBI Network), and LBA (the Australian Long Baseline Array) observations 
\citep{2011AJ....142...35P,2011AJ....142..105P,2011MNRAS.414.2528P,2012MNRAS.419.1097P,2013AJ....146....5P,2015ApJS..217....4S,2017ApJS..230...13S,2019MNRAS.485...88P}.
Their positions are determined and presented within the VLBI-based Radio Fundamental Catalogue\footnote{\url{http://astrogeo.org/vlbi/solutions/rfc_2019c/}} (RFC).
We note that a special effort was made by the VCS program observations to compile a complete sub-sample of AGN limited by the flux density integrated over VLBI images $S^\mathrm{VLBI}_\mathrm{8GHz}>150$~mJy at 8~GHz, and a similar effort was made with LBA observations. This complete sample consists of 3388 objects. The resulting sky coverage is shown with grey dots in \autoref{f:skymap}.

Note that the image database and the catalog contain the data for other wavelengths (2.3, 5, 15, 22~GHz) as well, and go down to lower flux density levels at 8~GHz. Altogether, the VLBI catalog contains the measurements for more than $16\,000$ AGN. However, the only deep statistically complete sample is the aforementioned one. Most of the other wavelengths lack the data below $-30\degr$ declination. The 15~GHz band is complete thanks to the \mbox{MOJAVE} project \citep{2019ApJ...874...43L} but only down to $S^{\text{VLBI}}_{\text{15GHz}} = 1.5$~Jy. 
 Generally, samples at different bands might be biased, e.g., towards $\gamma$-ray selected AGN \citep[e.g.,][]{2015ApJS..217....4S,2018ApJS..234...12L}, 
AGN seen through the galactic plane \citep{2011AJ....142...35P,2012MNRAS.419.1097P},
or optically bright AGN \citep{2011AJ....142..105P,2013AJ....146....5P}.
The 22~GHz sample might be biased towards the most compact AGN selected to serve for the high-frequency realization of the celestial reference frame \citep{2010AJ....139.1713C}. 
That is why, to achieve the most robust results, we use only the 8~GHz sample in our statistical studies.

In our analysis, we use the flux density integrated over VLBI images of AGN and call it throughout the paper the ``VLBI flux density.'' For most of the Doppler-boosted AGN that comprise our sample, it is dominated by emission of the apparent parsec-scale jet base, see our detailed discussion in \autoref{s:discussion_core}.
For the objects imaged by VLBI at more than one epoch, the average of all the measurements is used in the analysis. The number of the observations per source ranges from 1 to more than 150, with a median of 5.
The average we use throughout this paper is the geometric mean (or, equivalently, the arithmetic mean of logarithms) because the range of flux densities can cover several orders of magnitude, and relative differences are important.

\subsection{RATAN-600 AGN Monitoring} 
\label{s:data_radio}

The Russian RATAN-600 radio telescope \citep{1979S&T....57..324K} of the Special Astrophysical Observatory has been monitoring at 1-22~GHz a sample of AGN selected on their VLBI flux density since late 1980s. The details of these observations, the data analysis, the observing sample, and the results can be found in \citet{1997BSAO...44...50K,1999A&AS..139..545K,2000PASJ...52.1027K,2002PASA...19...83K}. The measurements of a target at a given observing epoch occur simultaneously at 1, 2, 5, 8, 11, and 22~GHz. For the analysis in this paper, we drop the lowest two frequencies since they are often affected by Radio Frequency Interference (RFI), which became stronger during the years used in this paper: 2009 -- 2019, inclusive.

The RATAN observing sample was originally selected on the basis of the correlated VLBI flux density measurements by \citet{1985AJ.....90.1599P} and was later supplemented with new objects found by the VCS survey. Thus, the sample contains AGN with strong parsec-scale radio jets and has good completeness characteristics down to $S^{\text{VLBI}}_{\text{8GHz}}\approx0.4$~Jy.
Due to the ring shape and the transit observing mode of the telescope, the best monitored part of the sample, with 3-4 epochs per year, is restricted to a declination range from $-30\degr$ to $+43\degr$. This range covers almost all of the IceCube high-energy track events in our sample. The full RATAN-600 dataset we use in our analysis has 1099 sources observed at least five times, 758 of which observed at least ten times.

There is a rich multi-frequency dataset produced by the F-GAMMA AGN broad-band spectrum monitoring program \citep{2016A&A...596A..45F}. Unfortunately, the published data cover the period until 2015 only \citep{2019A&A...626A..60A}. This is not long enough for our analysis since many neutrino events were collected after 2015. We have not used these data in the paper.

\section{Statistical analysis} 
\label{s:analysis}

\begin{figure}[t]
\includegraphics[width=1.0\columnwidth]{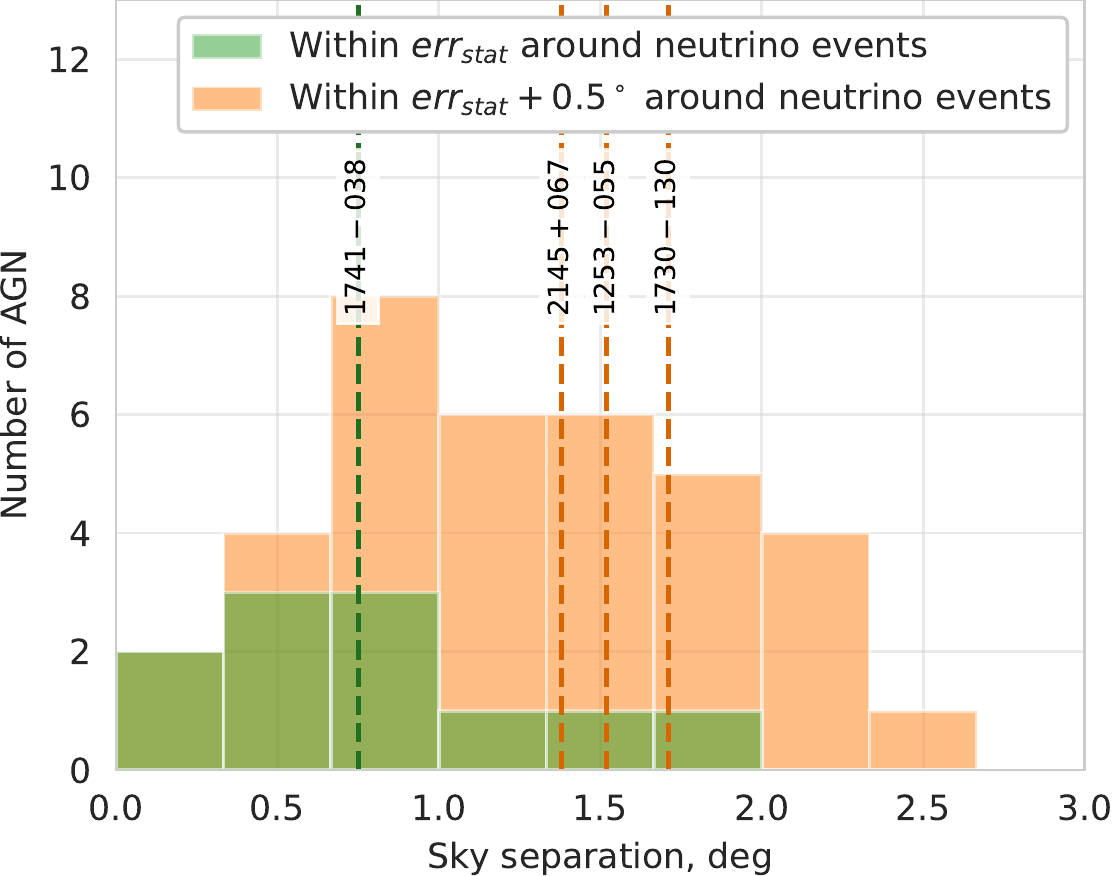}
\caption{
Distribution of angular distances between AGN and the corresponding neutrino events. Color differentiates AGN inside the IceCube statistical error regions (green, 11 objects) and those inside the regions enlarged to account for unknown systematic errors (orange, 36 objects), see \autoref{s:an_avgflux}. Vertical lines represent the four strongest AGN distinguished by our analysis (see \autoref{f:p_val_vs_count}).
\label{f:agn_dist_hist}
}
\end{figure}

\subsection{Flux Density of AGN Radio Emission from Parsec Scales}
\label{s:an_avgflux}

We use the average historic VLBI flux density of AGN (\autoref{s:data_vlbi}) to determine whether neutrino-emitting ones tend to be stronger in terms of their radio emission from compact parsec-scale central regions. 
We average the flux density over all the sources lying inside the error regions of IceCube events and take this value as the test statistic $v$. Then we test if it is significantly higher than could arise by chance for randomly-selected AGN. A Monte-Carlo method is employed for this testing in the following way:
\begin{itemize}
    \item Compute the statistic of interest using real positions of IceCube events. Denote its value as $v_\mathrm{real}$.
    \item Repeat $N=10000$ times the following:
    \begin{itemize}
        \item Shift IceCube events to random right ascension coordinates, keeping declinations and error regions unchanged\footnote{For the South Pole location of IceCube, this is equivalent to randomizing the sidereal arrival time of the event. To a good approximation, the sensitivity of the experiment depends on the zenith angle only \citep{IceCube-exposure}, and constant zenith angles correspond to constant declinations. Note certain drawbacks of this method for the cases when reshuffled error ellipses overlap with original ones, especially close to Celestial poles; however, for our purposes, this would result in a conservative estimate of the chance of random coincidence because any possible true correlation would only increase the background estimated in this way.};
        \item Compute the same statistic for these randomly shifted events in place of real ones. Denote this value $v_i, \quad 1 \leq i \leq N$.
    \end{itemize}
    \item The empirical distribution of $v_i$ represents the test statistic distribution under the null hypothesis that the statistic is not related to detected neutrinos. We compute confidence intervals for the null, which are shown in our plots later, using the quantiles of this distribution.
    \item Count random realizations with values not lower than the real one: $M = \sum_i [v_i \geq v_\mathrm{real}]$ (flip the sign to test the difference in the opposite direction). Calculate the $p$-value, defined as the probability of a chance coincidence, as \\$p = \displaystyle\frac{M+1}{N+1}$ following \citet{Davison2013BMA2556084}.
\end{itemize}

To implement this procedure, we need to specify error regions for each event. We start with 90\% coordinate-wise statistical uncertainties in Right Ascension and Declination reported for IceCube events and transform them to obtain two-dimensional 90\% coverage regions. Specifically, we multiply the coordinate-wise errors by the ratio of 90\% quantiles of two- and one-dimensional Gaussian distributions: $\displaystyle\frac{\sqrt{-\log{(1-0.9)}}}{\mathrm{erf}^{-1}(0.9)} \approx 1.30$. This leads to regions bounded by four quarters of ellipses, as IceCube reports two-sided uncertainties for each coordinate.

Next we need to account for systematic errors in IceCube event positions. As mentioned in \autoref{s:data_icecube}, these errors are always present but their values are not published. Thus, we choose to introduce the systematic error magnitude as a free parameter --- same for all events and directions on the sky --- and determine its optimal value. This is implemented by a procedure commonly used in particle and astroparticle physics, \citep[see, e.g., for its application to cosmic-ray arrival directions][]{Tinyakov:2003bi}. The procedure consists of trying multiple values of the unknown parameter to select one with the strongest signal. If done naively, this is affected by the \emph{multiple comparisons issue}: for $K$ trials, one expects to obtain a $p$-value as low as $1/K$ at least once just by a statistical fluctuation. Thus, to ensure an unbiased \emph{post-trial} result a correction is needed. We use a Monte-Carlo procedure to account for multiple trials in fitting the unknown parameter. First, the \emph{pre-trial} $p$-value is computed as follows:
\begin{itemize}
    \item For each assumed value of the systematic error $0 \leq x \leq 1\degr$ (we take 11 values spaced by $0.1\degr$) compute the raw $p$-value as described at the beginning of this section. The only difference is that the error regions of all events are increased by linearly adding $x$ in all directions. Denote these $p$-values as $p_j$, $j=1,\dots 11$.
    \item Take the minimum of those $p_j$ values, which corresponds to the value of $x$ giving the most significant flux density difference. This minimum is called the pre-trial $p$-value.
\end{itemize}
The final post-trial $p$-value is calculated by repeating these steps for artificial Monte-Carlo samples to determine how often they yield a lower $p$ (a more significant difference) by a chance fluctuation. This approach is equivalent to plugging the pre-trial $p$-value as the test statistic $v$ into the Monte-Carlo testing method outlined above. The computed post-trial $p$-value is thus unaffected by the multiple comparisons issue.

\begin{figure}[t]
\includegraphics[width=1.0\columnwidth]{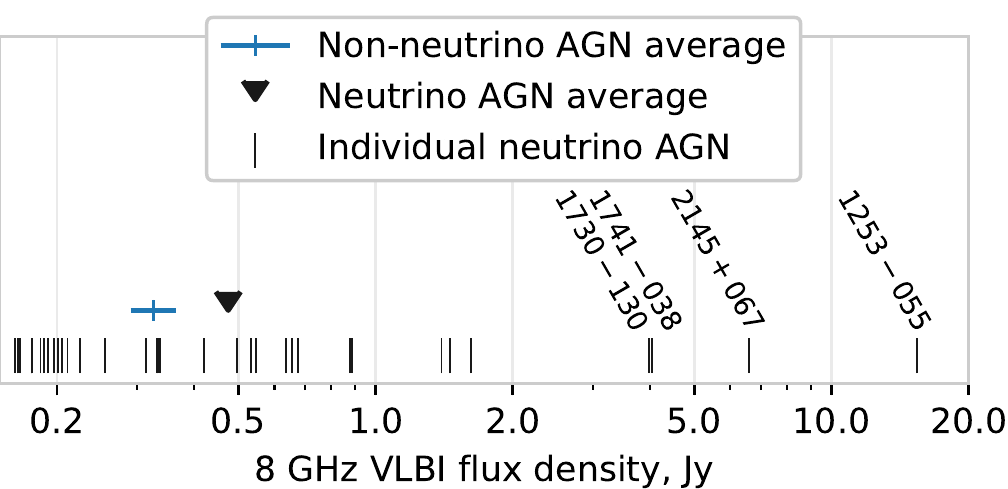}
\caption{
Average of VLBI flux densities for AGN inside the IceCube error regions shown as a black triangle in comparison to 68\% Monte-Carlo interval (blue horizontal line) for randomly-shifted events. Flux densities for individual AGN inside the error regions are also shown as vertical black ticks for information.
\label{f:avg_fluxes}
}
\end{figure}

This approach results in the chance probability $p=0.2\%$ of the average flux density of AGN around IceCube detections being as high as observed; thus, we conclude that the effect is significant. The minimum pre-trial $p$-value is $0.09\%$ obtained for the additional error of $x = 0.5\degr$. This $x$ can be interpreted as a rough estimate of IceCube systematic errors, though more knowledge about the distribution of statistical uncertainty than available in the event catalogs is required to study it in more detail. We note that our result is in a very good agreement with the independent IceCube systematic errors estimate, $<1\degr$, discussed in \autoref{s:data_icecube}. Further in this subsection and in Figures \ref{f:skymap}, \ref{f:agn_dist_hist}, and \ref{f:avg_fluxes}, we use the statistical error regions enlarged by this value, $x=0.5\degr$.

\autoref{f:skymap} demonstrates IceCube events on the sky together with AGN from our complete sample. \autoref{f:agn_dist_hist} specifically illustrates changes in the number of AGN and in the angular distance distribution when taking systematic errors into account. \autoref{f:avg_fluxes} compares the average of actual VLBI flux density values for AGN within the neutrino error regions to Monte-Carlo realizations of this average for randomly-shifted positions of neutrino events. This figure highlights that the actual AGN being selected as possible neutrino counterparts are, on average, stronger on parsec scales. 
Note that the same analysis we performed for AGN observations at 2, 5, 15, and 22~GHz resulted in a similar outcome. However, we do not use these results here because only the 8~GHz VLBI sample has the desired completeness as discussed in \autoref{s:data_vlbi}. 

We stress that VLBI observations are crucial for this result. This can be illustrated by repeating the same analysis for the NVSS \citep[NRAO VLA Sky Survey,][]{Condon_1998} catalog containing a complete sample of 2 million radio sources \emph{without} selection by the compact VLBI component. We find that it does not show any significant difference in flux density between the sources inside IceCube error regions and randomly selected ones. However, limiting this analysis to the intersection of NVSS and our 8 GHz VLBI complete sample (2919 sources) leads to a marginally significant difference in NVSS flux density: minimum pre-trial $p$-value is 2\%. This effect does not appear when analysing the same number of sources selected as strongest by NVSS flux density itself. It would be interesting to analyze VLASS \citep[VLA Sky Survey,][]{2018AAS...23123108M} in this way when the data become available, as it has higher sensitivity and resolution compared to NVSS, and probes scales closer to those of VLBI.

\begin{figure}
\includegraphics[width=1.0\columnwidth]{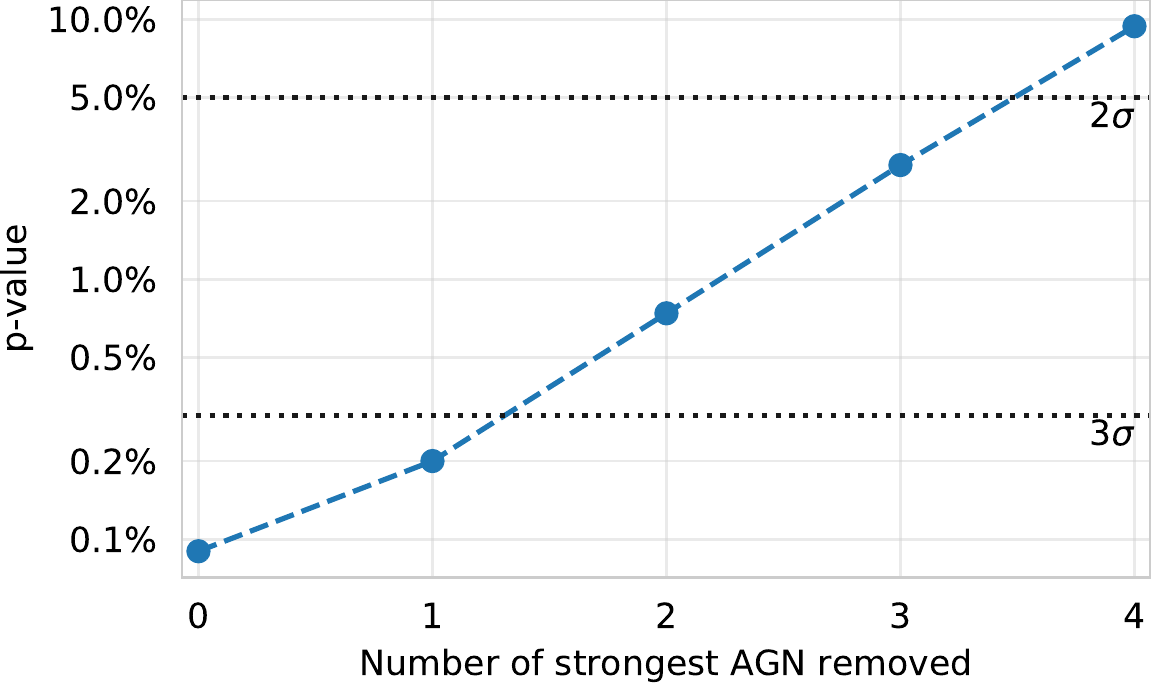}
\caption{
Significance level of AGN within IceCube error regions being stronger in terms of VLBI flux density when removing up to four strongest sources from the analysis. Horizontal lines indicate significance levels corresponding to $2\sigma$ and $3\sigma$ difference for a Gaussian distribution.
\label{f:p_val_vs_count}
}
\end{figure}

Now, after we have established that neutrino-emitting AGN have stronger compact radio emission than average, the next logical step is to estimate how many sources drive this effect. We repeat our analysis dropping the strongest sources in terms of their flux density one by one until the significance disappears, as illustrated in \autoref{f:p_val_vs_count}. The $p$-value rises above 5\% level when four objects are removed, and we interpret this as a lower bound on the number of AGN likely emitting high-energy neutrinos. The four strongest sources are 
\object{1253$-$055} (\object{3C\,279}), \object{PKS~2145$+$067}, \object{PKS~1741$-$038}, and \object{1730$-$130} (\object{NRAO~530}). See \autoref{t:associations} for their properties. None of these AGN has been singled out as sources of the observed IceCube neutrinos in the literature before. We show their names in all the plots containing individual sources: Figures \ref{f:skymap}, \ref{f:agn_dist_hist}, and \ref{f:avg_fluxes}

Note that the \object{TXS~0506+056} blazar possibly associated with neutrino detection 170922A \citep{IceCubeTXSgamma} is not among those four AGN.
Its average VLBI flux density from 13 observing epochs in 1995-2018 is only 0.4~Jy, not much higher than the average in the whole sample. However, its single-dish and VLBI flux density rose up to more than 1.5~Jy by 2019 \citep[e.g.,][see also public MOJAVE 15~GHz VLBA data\footnote{\url{http://www.physics.purdue.edu/astro/MOJAVE/sourcepages/0506+056.shtml}}]{r:ros0506,r:kovalev0506}.
Another notable AGN not included in these four strongest AGN is the quasar PKS~1502+106 \citep[e.g.,][]{r:Fermi1502} directionally coincident to a recent IceCube event 190730A \citep{2019ATel12967....1T}. Its average VLBI flux density from 17 epochs in 2001-2018 is 1.5~Jy; whereas its flux density rose in 2019 according to the Ovens Valley Radio Observatory \citep[OVRO,][]{r:kielmann1502ATel}, MOJAVE observations\footnote{\url{http://www.physics.purdue.edu/astro/MOJAVE/sourcepages/1502+106.shtml}}, and RATAN-600 (\autoref{t:associations_ratan}) to the level of 3-4~Jy.
This suggests that the four brightest AGN listed above as the most probable neutrino associations do not exhaust neutrino sources in the VLBI-selected AGN list. This may partly be due to the historic average VLBI flux density values being used. Evidently, the next step should be a temporal correlation analysis.

\begin{table*}
\caption{IceCube high-energy neutrino events positionally associated with VLBI-compact AGN
\label{t:associations}
}
\begin{center}
\begin{tabular}{ccc|cccrrrr}
\hline\hline
\multicolumn{3}{c|}{IceCube event} & \multicolumn{7}{c}{AGN} \\
Date & Category & $E$ & \multicolumn{2}{c}{Name} & $z$ & $S^{\text{VLBI}}_{\text{8GHz}}$ & $d$ & $d - \mathrm{err}_{\text{event}}$ & $\gamma$-ray Flux \\
 & & (TeV) & B1950 & J2000 & & (Jy) & ($\degr$) & ($\degr$) & ($10^{-9}\mathrm{cm}^{-2} \mathrm{s}^{-1}$) \\
(1) & (2) & (3) & (4) & (5) & (6) & (7) & (8) & (9) & (10) \\
\hline
\multirow{2}{*}{2010-10-09}  & \multirow{2}{*}{EHEA}        & \multirow{2}{*}{660}         & \texttt{\object{2201+098}} & \texttt{\object{J2203+1007}} & 1.00                       & 0.20                       & 0.99                       & 0.23                       & \n                         \\
                             &                              &                              & \texttt{\object{2157+102}} & \texttt{\object{J2200+1030}} & \n                         & 0.19                       & 1.20                       & 0.33                       & 0.24                       \\
\hline
\multirow{2}{*}{2010-11-13}  & \multirow{2}{*}{MUONT}       & \multirow{2}{*}{520}         & \texttt{\object{1855+031}} & \texttt{\object{J1858+0313}} & \n                         & 0.68                       & 1.44                       & 0.00                       & \n                         \\
                             &                              &                              & \texttt{\object{1853+027}} & \texttt{\object{J1855+0251}} & \n                         & 0.16                       & 2.07                       & 0.19                       & \n                         \\
\hline
\multirow{1}{*}{2011-07-14}  & \multirow{1}{*}{HESEA}       & \multirow{1}{*}{253}         & \texttt{\object{0429+415}} & \texttt{\object{J0432+4138}} & 1.02                       & 1.39                       & 1.34                       & 0.41                       & \n                         \\
\hline
\multirow{1}{*}{2011-09-30}  & \multirow{1}{*}{EHEA}        & \multirow{1}{*}{\n }         & \textbf{\texttt{\object{1741-038}}} & \textbf{\texttt{\object{J1743-0350}}} & \textbf{1.05}              & \textbf{4.04}              & \textbf{0.75}              & \textbf{0.00}              & \textbf{0.36}              \\
\hline
\multirow{1}{*}{2012-05-23}  & \multirow{1}{*}{EHEA}        & \multirow{1}{*}{\n }         & \texttt{\object{1123+264}} & \texttt{\object{J1125+2610}} & 2.35                       & 0.88                       & 0.44                       & 0.00                       & \n                         \\
\hline
\multirow{1}{*}{2012-09-22}  & \multirow{1}{*}{EHEA}        & \multirow{1}{*}{\n }         & \texttt{\object{0435+217}} & \texttt{\object{J0438+2153}} & 1.30                       & 0.20                       & 2.30                       & 0.49                       & 0.30                       \\
\hline
\multirow{1}{*}{2012-10-11}  & \multirow{1}{*}{EHEA}        & \multirow{1}{*}{210}         & \texttt{\object{1337-013}} & \texttt{\object{J1340-0137}} & 1.62                       & 0.21                       & 0.79                       & 0.11                       & 0.33                       \\
\hline
\multirow{1}{*}{2013-06-27}  & \multirow{1}{*}{HESEA}       & \multirow{1}{*}{200}         & \texttt{\object{0611+131}} & \texttt{\object{J0613+1306}} & 0.74                       & 0.33                       & 0.91                       & 0.00                       & \n                         \\
\hline
\multirow{1}{*}{2013-10-14}  & \multirow{1}{*}{MUONT}       & \multirow{1}{*}{390}         & \texttt{\object{0208+106}} & \texttt{\object{J0211+1051}} & 0.20                       & 0.66                       & 0.67                       & 0.22                       & 5.32                       \\
\hline
\multirow{1}{*}{2013-10-23}  & \multirow{1}{*}{EHEA}        & \multirow{1}{*}{\n }         & \texttt{\object{2007+131}} & \texttt{\object{J2009+1318}} & \n                         & 0.18                       & 1.89                       & 0.36                       & \n                         \\
\hline
\multirow{2}{*}{2013-12-04}  & \multirow{2}{*}{EHEA}        & \multirow{2}{*}{\n }         & \texttt{\object{1909-151}} & \texttt{\object{J1912-1504}} & \n                         & 0.31                       & 1.36                       & 0.23                       & \n                         \\
                             &                              &                              & \texttt{\object{1914-154}} & \texttt{\object{J1916-1519}} & \n                         & 0.18                       & 1.07                       & 0.01                       & 0.36                       \\
\hline
\multirow{1}{*}{2014-01-08}  & \multirow{1}{*}{EHEA}        & \multirow{1}{*}{\n }         & \texttt{\object{2256+017}} & \texttt{\object{J2258+0203}} & 2.66                       & 0.18                       & 0.53                       & 0.05                       & \n                         \\
\hline
\multirow{1}{*}{2014-02-03}  & \multirow{1}{*}{EHEA}        & \multirow{1}{*}{\n }         & \texttt{\object{2325-150}} & \texttt{\object{J2327-1447}} & 2.46                       & 0.53                       & 2.58                       & 0.08                       & \n                         \\
\hline
\multirow{1}{*}{2015-01-27}  & \multirow{1}{*}{MUONT}       & \multirow{1}{*}{210}         & \texttt{\object{0643+057}} & \texttt{\object{J0645+0541}} & \n                         & 0.16                       & 1.48                       & 0.48                       & \n                         \\
\hline
\multirow{3}{*}{2015-08-12}  & \multirow{3}{*}{EHEA}        & \multirow{3}{*}{380}         & \textbf{\texttt{\object{2145+067}}} & \textbf{\texttt{\object{J2148+0657}}} & \textbf{0.99}              & \textbf{6.60}              & \textbf{1.38}              & \textbf{0.41}              & \textbf{0.22}              \\
                             &                              &                              & \texttt{\object{2149+069}} & \texttt{\object{J2151+0709}} & 1.36                       & 0.89                       & 1.00                       & 0.39                       & \n                         \\
                             &                              &                              & \texttt{\object{2149+056}} & \texttt{\object{J2151+0552}} & 0.74                       & 0.50                       & 0.44                       & 0.00                       & \n                         \\
\hline
\multirow{1}{*}{2015-08-31}  & \multirow{1}{*}{EHEA}        & \multirow{1}{*}{\n }         & \texttt{\object{0333+321}} & \texttt{\object{J0336+3218}} & 1.26                       & 1.62                       & 1.76                       & 0.29                       & 0.46                       \\
\hline
\multirow{1}{*}{2015-09-04}  & \multirow{1}{*}{MUONT}       & \multirow{1}{*}{220}         & \texttt{\object{0849+287}} & \texttt{\object{J0852+2833}} & 1.28                       & 0.33                       & 1.03                       & 0.38                       & 0.19                       \\
\hline
\multirow{1}{*}{2015-09-26}  & \multirow{1}{*}{EHEA}        & \multirow{1}{*}{\n }         & \textbf{\texttt{\object{1253-055}}} & \textbf{\texttt{\object{J1256-0547}}} & \textbf{0.54}              & \textbf{15.38}             & \textbf{1.52}              & \textbf{0.26}              & \textbf{24.53}             \\
\hline
\multirow{1}{*}{2015-11-14}  & \multirow{1}{*}{MUONT}       & \multirow{1}{*}{740}         & \texttt{\object{0459+135}} & \texttt{\object{J0502+1338}} & 0.45                       & 0.64                       & 1.22                       & 0.39                       & 0.31                       \\
\hline
\multirow{3}{*}{2016-01-28}  & \multirow{3}{*}{EHEA}        & \multirow{3}{*}{\n }         & \textbf{\texttt{\object{1730-130}}} & \textbf{\texttt{\object{J1733-1304}}} & \textbf{0.90}              & \textbf{3.98}              & \textbf{1.71}              & \textbf{0.43}              & \textbf{6.66}              \\
                             &                              &                              & \texttt{\object{1735-150}} & \texttt{\object{J1738-1503}} & \n                         & 0.18                       & 1.14                       & 0.00                       & \n                         \\
                             &                              &                              & \texttt{\object{1739-152}} & \texttt{\object{J1742-1517}} & \n                         & 0.17                       & 2.13                       & 0.47                       & 0.39                       \\
\hline
\multirow{1}{*}{2016-03-31}  & \multirow{1}{*}{MUONT}       & \multirow{1}{*}{380}         & \texttt{\object{0103+156}} & \texttt{\object{J0105+1553}} & \n                         & 0.20                       & 0.88                       & 0.31                       & \n                         \\
\hline
\multirow{2}{*}{2017-03-21}  & \multirow{2}{*}{EHEA}        & \multirow{2}{*}{\n }         & \texttt{\object{0629-141}} & \texttt{\object{J0631-1410}} & 1.02                       & 0.55                       & 0.96                       & 0.00                       & \n                         \\
                             &                              &                              & \texttt{\object{0628-133}} & \texttt{\object{J0630-1323}} & 1.02                       & 0.34                       & 1.72                       & 0.17                       & \n                         \\
\hline
\multirow{1}{*}{2017-09-22}  & \multirow{1}{*}{EHEA}        & \multirow{1}{*}{290}         & \texttt{\object{0506+056}} & \texttt{\object{J0509+0541}} & 0.34                       & 0.42                       & 0.08                       & 0.00                       & 5.99                       \\
\hline
\multirow{1}{*}{2017-11-06}  & \multirow{1}{*}{EHEA}        & \multirow{1}{*}{\n }         & \texttt{\object{2235+071}} & \texttt{\object{J2238+0724}} & 1.01                       & 0.22                       & 0.45                       & 0.00                       & \n                         \\
\hline
\multirow{1}{*}{2018-09-08}  & \multirow{1}{*}{EHEA}        & \multirow{1}{*}{\n }         & \texttt{\object{0943-016}} & \texttt{\object{J0945-0153}} & 2.37                       & 0.26                       & 1.87                       & 0.00                       & \n                         \\
\hline
\multirow{3}{*}{2019-07-30}  & \multirow{3}{*}{GOLD}        & \multirow{3}{*}{299}         & \texttt{\object{1502+106}} & \texttt{\object{J1504+1029}} & 1.84                       & 1.46                       & 0.31                       & 0.00                       & 19.01                      \\
                             &                              &                              & \texttt{\object{1451+106}} & \texttt{\object{J1453+1025}} & 1.77                       & 0.19                       & 2.32                       & 0.49                       & \n                         \\
                             &                              &                              & \texttt{\object{1500+094}} & \texttt{\object{J1503+0917}} & \n                         & 0.18                       & 1.17                       & 0.01                       & \n                         \\
\hline
\hline
\end{tabular}
\end{center}
\tablecomments{
AGN from the complete VLBI sample that fall into the error regions of IceCube high-energy neutrino detections assuming the systematic error of $0.5\degr$. See \autoref{s:data_vlbi} for the characteristics of the VLBI sample, \autoref{s:data_icecube} for IceCube events selection criteria and category notations, and \autoref{s:an_avgflux} for details on how we estimate the systematic uncertainty. If several AGN from our sample are found within the error region of a given event, we list them all in order of decreasing VLBI flux density.
The four most probable neutrino associations from the analysis of historic average VLBI flux densities are shown by the boldface font, see \autoref{s:an_avgflux} for details.
Columns are as follows: 
(1), (2), (3) IceCube event parameters; (4), (5) AGN name in B1950 and J2000 formats; (6) Redshift taken from NASA/IPAC Extragalactic Database; (7) Average VLBI flux density at 8 GHz; (8) Angular separation on the sky between the AGN and corresponding IceCube event; (9) Angular separation between the AGN and the $90\%$ statistical uncertainty ellipse of the corresponding IceCube event; (10) Gamma ray (1-100 GeV) flux of the AGN as measured by \textit{Fermi} LAT \citep{r:4fgl_cat}.
}
\end{table*}

\begin{table*}
\caption{
AGN within IceCube events error regions monitored by RATAN-600
\label{t:associations_ratan}
}
\begin{center}
\begin{tabular}{ccc|llcp{0.9cm}rrrrr}
\hline\hline
\multicolumn{3}{c|}{IceCube event} & \multicolumn{9}{c}{AGN} \\
Date & Category & $E$ & \multicolumn{2}{c}{Name} & $z$ & \#\enspace of Epochs  & $S^{\text{RATAN}}_{\text{22GHz}}$ & $R_{\text{22GHz}}^{t=0}$ & $d$ & $d - \mathrm{err}_{\text{event}}$ & $\gamma$-ray Flux \\
 & & (TeV) & B1950 & J2000 & & & (Jy) & & ($^\circ$) & ($^\circ$) & ($10^{-9}\mathrm{cm}^{-2} \mathrm{s}^{-1}$) \\
(1) & (2) & (3) & (4) & (5) & (6) & (7) & (8) & (9) & (10) & (11) & (12) \\
\hline
\multirow{2}{*}{2011-07-14}  & \multirow{2}{*}{HESEA}       & \multirow{2}{*}{253}         & \texttt{\object{0429+415}} & \texttt{\object{J0432+4138}} & 1.02 & 38 & 1.11 & 1.06 & 1.34 & 0.41 & \n  \\
                             &                              &                              & \tablenotemark{a}\texttt{\object{0424+414}} & \texttt{\object{J0427+4133}} & \n  & 8  & 0.37 & 1.22 & 1.41 & 0.53 & \n  \\
\hline
\multirow{1}{*}{2011-09-30}  & \multirow{1}{*}{EHEA}        & \multirow{1}{*}{\n }         & \texttt{\object{1741-038}} & \texttt{\object{J1743-0350}} & 1.05 & 38 & 3.40 & 1.37 & 0.75 & 0.00 & 0.36 \\
\hline
\multirow{1}{*}{2012-05-23}  & \multirow{1}{*}{EHEA}        & \multirow{1}{*}{\n }         & \texttt{\object{1123+264}} & \texttt{\object{J1125+2610}} & 2.35 & 32 & 0.58 & 1.26 & 0.44 & 0.00 & \n  \\
\hline
\multirow{1}{*}{2013-06-27}  & \multirow{1}{*}{HESEA}       & \multirow{1}{*}{200}         & \texttt{\object{0611+131}} & \texttt{\object{J0613+1306}} & 0.74 & 17 & 0.31 & 1.42 & 0.91 & 0.00 & \n  \\
\hline
\multirow{1}{*}{2014-01-08}  & \multirow{1}{*}{EHEA}        & \multirow{1}{*}{\n }         & \texttt{\object{2256+017}} & \texttt{\object{J2258+0203}} & 2.66 & 24 & 0.25 & 2.01 & 0.53 & 0.05 & \n  \\
\hline
\multirow{1}{*}{2014-02-03}  & \multirow{1}{*}{EHEA}        & \multirow{1}{*}{\n }         & \texttt{\object{2325-150}} & \texttt{\object{J2327-1447}} & 2.46 & 25 & 0.40 & 1.92 & 2.58 & 0.08 & \n  \\
\hline
\multirow{3}{*}{2015-08-12}  & \multirow{3}{*}{EHEA}        & \multirow{3}{*}{380}         & \texttt{\object{2145+067}} & \texttt{\object{J2148+0657}} & 0.99 & 40 & 2.55 & 1.16 & 1.38 & 0.41 & 0.22 \\
                             &                              &                              & \texttt{\object{2149+069}} & \texttt{\object{J2151+0709}} & 1.36 & 27 & 0.55 & 0.93 & 1.00 & 0.39 & \n  \\
                             &                              &                              & \texttt{\object{2149+056}} & \texttt{\object{J2151+0552}} & 0.74 & 27 & 0.31 & 0.92 & 0.44 & 0.00 & \n  \\
\hline
\multirow{1}{*}{2015-08-31}  & \multirow{1}{*}{EHEA}        & \multirow{1}{*}{\n }         & \texttt{\object{0333+321}} & \texttt{\object{J0336+3218}} & 1.26 & 41 & 1.54 & 0.63 & 1.76 & 0.29 & 0.46 \\
\hline
\multirow{1}{*}{2015-09-26}  & \multirow{1}{*}{EHEA}        & \multirow{1}{*}{\n }         & \texttt{\object{1253-055}} & \texttt{\object{J1256-0547}} & 0.54 & 42 & 17.10 & 1.01 & 1.52 & 0.26 & 24.53 \\
\hline
\multirow{1}{*}{2015-11-14}  & \multirow{1}{*}{MUONT}       & \multirow{1}{*}{740}         & \texttt{\object{0459+135}} & \texttt{\object{J0502+1338}} & 0.45 & 32 & 0.41 & 1.32 & 1.22 & 0.39 & 0.31 \\
\hline
\multirow{1}{*}{2016-01-28}  & \multirow{1}{*}{EHEA}        & \multirow{1}{*}{\n }         & \texttt{\object{1730-130}} & \texttt{\object{J1733-1304}} & 0.90 & 41 & 3.68 & 0.84 & 1.71 & 0.43 & 6.66 \\
\hline
\multirow{2}{*}{2017-09-22}  & \multirow{2}{*}{EHEA}        & \multirow{2}{*}{290}         & \tablenotemark{a}\texttt{\object{0502+049}} & \texttt{\object{J0505+0459}} & 0.95 & 26 & 0.65 & 1.34 & 1.30 & 0.63 & 6.56 \\
                             &                              &                              & \texttt{\object{0506+056}} & \texttt{\object{J0509+0541}} & 0.34 & 36 & 0.45 & 1.59 & 0.08 & 0.00 & 5.99 \\
\hline
\multirow{1}{*}{2018-10-23}  & \multirow{1}{*}{EHEA}        & \multirow{1}{*}{\n }         & \tablenotemark{a}\texttt{\object{1749-101}} & \texttt{\object{J1752-1011}} & \n  & 29 & 0.29 & 1.68 & 2.58 & 0.62 & 0.49 \\
\hline
\multirow{1}{*}{2019-07-30}  & \multirow{1}{*}{GOLD}        & \multirow{1}{*}{299}         & \texttt{\object{1502+106}} & \texttt{\object{J1504+1029}} & 1.84 & 35 & 1.45 & 3.14 & 0.31 & 0.00 & 19.01 \\
\hline
\hline
\end{tabular}
\end{center}
\tablecomments{
AGN from the RATAN-600 monitoring program which fall into the error regions of IceCube high-energy neutrino detections assuming the systematic error of $0.7\degr$. See \autoref{s:data_radio} for the monitoring program characteristics, \autoref{s:data_icecube} for IceCube events selection criteria and category notations, and \autoref{s:an_variability} for details on how we estimate the systematic uncertainty. If several AGN from the monitoring program are found within an error region of a given event, we list them all ordered by decreasing average flux density.\\
Columns are as follows: 
(1) --- (6) and (10) --- (12) Same as corresponding columns in \autoref{t:associations}; (7) Number of observations by RATAN-600 from 2009 to 2019; (8) Average flux density at 22 GHz; (9) Ratio of the average flux density within 0.9~yr (i.e., $\pm 0.5$~yr) around the corresponding IceCube event to the average outside of this period.
}
\tablenotetext{a}{Sources absent in \autoref{t:associations} due to slightly different assumed systematic errors.}
\end{table*}

\subsection{Temporal Correlations of Radio and Neutrino Observations}
\label{s:an_variability}

It is expected \citep[see, e.g.,][]{Murase-rev} that neutrinos can be associated with flares in central regions of AGN --- the immediate vicinity of the black hole or parts of the jet close to its origin. The studies of TXS 0506+056 \citep[e.g.,][]{IceCubeTXSgamma,r:kovalev0506,r:ros0506} support this prediction, however it has not been confirmed yet for larger samples of AGN. We approach the problem of associating neutrino to flares with a search for an excess in the radio flux density from AGN in temporal coincidence with IceCube neutrino events.

\begin{figure*}[tb]
\gridline{\fig{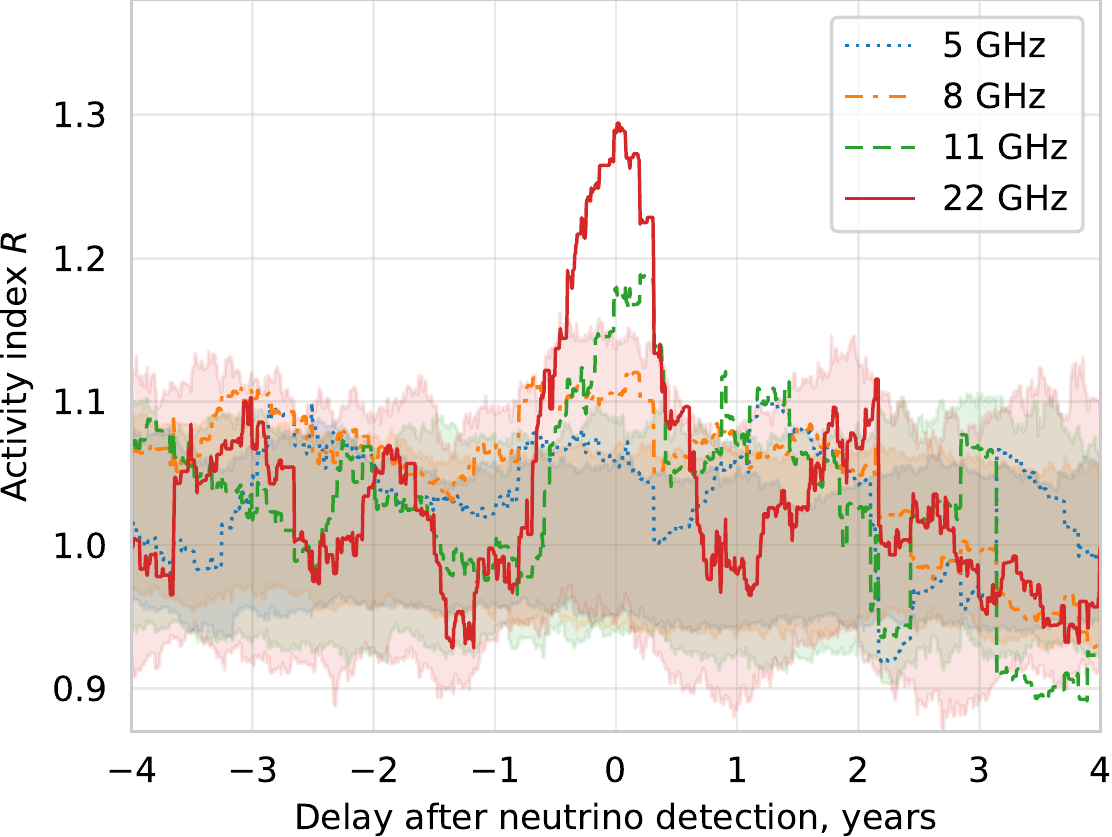}{0.5\textwidth}{(a) All sources: 18~AGN close to neutrino events.}
          \fig{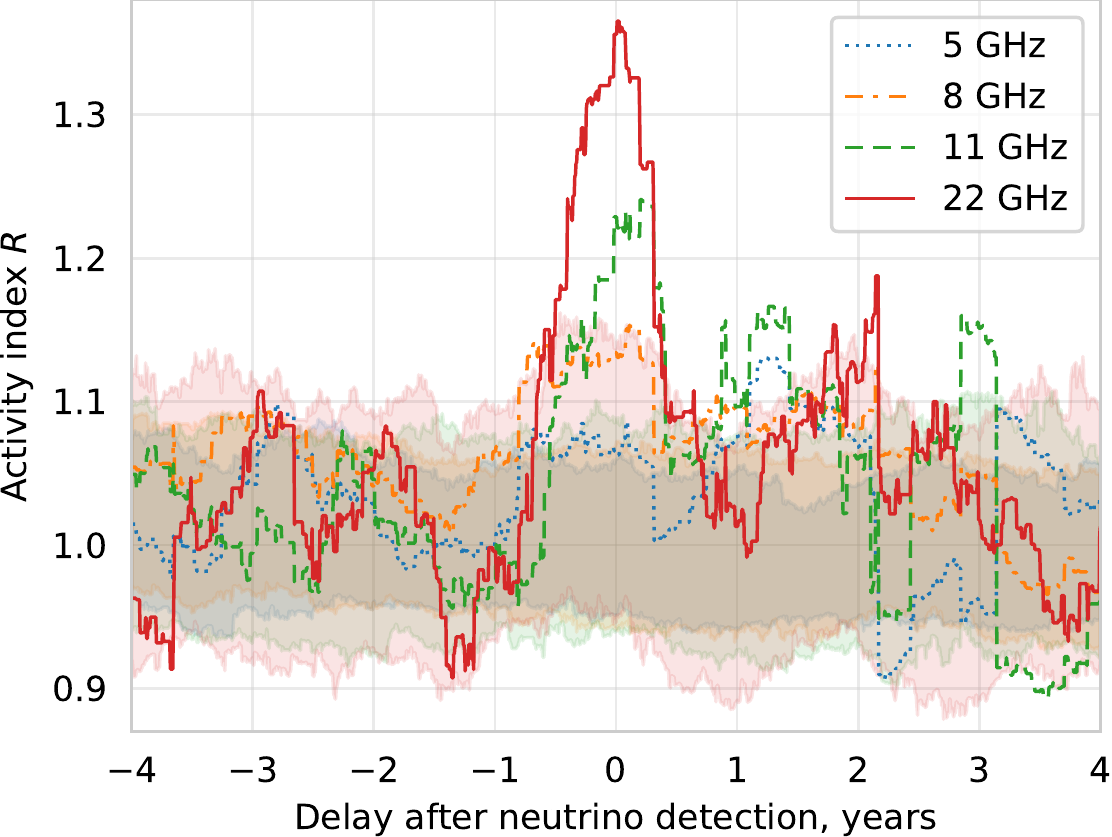}{0.5\textwidth}{(b) Without the four strongest sources (see \autoref{t:associations}): \object{1253$-$055}, \object{1730$-$130}, \object{1741$-$038}, and \object{2145$+$067}.}
          }
\caption{Ratio of RATAN-600 flux densities averaged over a 0.9~yr window to the average flux density outside it. Each point of the curve represents this ratio averaged across all AGN inside neutrino error regions versus the time delay between a 0.9~yr window center and the corresponding IceCube event. Filled areas correspond to curves of the same color and indicate pointwise 68\% intervals of Monte-Carlo realizations for randomly shifted neutrino event positions.
\label{f:xcorr}
}
\end{figure*}

We study the correlation of radio variability with neutrino detections by employing the RATAN-600 AGN monitoring data covering the time range 2009-2019, inclusive (\autoref{s:data_radio}). The dataset does not depend in any way on the VLBI measurements used in the previous subsection, so the following analysis represents an independent test of AGN being sources of $E>200$~TeV neutrinos. We chose to use the observations at the highest RATAN frequency of 22~GHz for our statistical analysis, as flares are typically more pronounced at shorter wavelengths due to synchrotron opacity effects, see \autoref{s:discussion_core} for details.

For each source within the IceCube error regions, we compute the \textit{radio activity index} $R_{\text{22GHz}}^{t=0}$ defined as the ratio of average RATAN flux density within a $\Delta T$ window (i.e., $\pm \Delta T / 2$) around neutrino detection to the average value outside of this time range.
Then the ratios $R_{\text{22GHz}}^{t=0}$ corresponding to all the sources within the error regions are averaged to form a single number --- the test statistic. This value being higher than can be expected from statistical fluctuations would mean that neutrinos do correlate with flares seen in radio observations. We test this hypothesis in the very same way as described in \autoref{s:an_avgflux}: plug $R_{\text{22GHz}}^{t=0}$ as the test statistic $v$ and use an additional trial range of $0.1\,\textrm{yr}\le\Delta T \le 2\,\textrm{yr}$ with 20 values spaced by $0.1\,\textrm{yr}$. The post-trial $p$-value is 5\%, which is not strongly significant, but, in the context of our results in \autoref{s:an_avgflux}, can definitely be considered suggestive. For comparison, the minimum raw pre-trial $p$-value is 1\% obtained at $\Delta T = 0.9$~yr and additional positional error of neutrino events equal to $0.7\degr$.  We list the ratio $R_{\text{22GHz}}^{t=0}$ for each AGN in \autoref{t:associations_ratan} using these values of $\Delta T$ and positional error. Note that the temporal correlation is less statistically significant than the average VLBI flux density difference analyzed in \autoref{s:an_avgflux}, which is expected: there are much fewer AGN in the RATAN monitoring program compared to the complete VLBI sample.

It is worth noting that the optimal value for the systematic error in \autoref{s:an_avgflux} was slightly different, $0.5\degr$. This is a perfect illustration of the statistical nature of our approach. Indeed, the analyses in \autoref{s:an_avgflux} and \autoref{s:an_variability} are based on completely different sets of radio data and, because of the smaller number of monitored sources, only a subset of our neutrino sample contributed to the second analysis. As we have already pointed out, see \autoref{s:data_icecube}, the IceCube systematic errors vary from event to event but we do not take this variation into account. Consequently, one expects a certain difference between average systematic errors for different sets of events. This is precisely what we observe: the values of additional errors determined in two analyses are different but are close to each other. This represents an additional consistency check for our analysis. Because of slightly larger error regions for the variability study, one neutrino event and three sources not present in \autoref{t:associations} contribute to the results of this subsection, see \autoref{t:associations_ratan}.

To visualize the correlation, we compute the activity index $R^t$ for different time lags: RATAN-600 measurements for all the sources are artificially shifted in time by $t$, whereas neutrino detection dates stay fixed. Note that the shifting is done for illustration only and is not used in the statistical analysis: $R^{t=0}$ already incorporates observations at all epochs in the form of averaging, and other $R^t$ are not independent of it. We show the ratio $R^t$ averaged across all AGN versus time lag for the four frequency bands of 5, 8, 11, and 22~GHz in \autoref{f:xcorr}. The values of $\Delta T$ and additional positional errors are those giving the lowest $p$-value for the zero-delay comparison, as explained above. This plot indicates that at the highest frequencies, 22~GHz and to a lesser extent 11 GHz, there is a pronounced peak around zero delay, whereas no structure is visible at 8 and 5 GHz. This is in a qualitative agreement with the nature of VLBI parsec-scale jet radio emission, see \autoref{s:discussion} for a detailed discussion.

As illustrated by \autoref{f:xcorr}(a), the correlation we detect happens on timescales of months. These are the smallest scales we are able to probe here due to the cadence of RATAN monitoring, so the question whether there is an even stronger correlation at the days or weeks timescale remains open. We stress that the RATAN-600 monitoring sample was originally selected based on VLBI observations, see \autoref{s:data_radio}. Extending this analysis to datasets of other monitoring programs with different selection criteria may thus require filtering their samples by the VLBI flux density.

The correlation found in \autoref{s:an_avgflux} is driven by four particular strongest sources indicated in \autoref{t:associations}. To make the test presented in this subsection independent of their contribution, we remove these four sources and repeat the analysis using the remaining 14 objects from \autoref{t:associations_ratan}. This does not reduce the significance of the temporal correlations, cf.\ \autoref{f:xcorr}(b).
Thus, we conclude that the temporal correlation is driven by additional AGN, not just those four distinguished by their time-averaged VLBI flux density in \autoref{s:an_avgflux}.
Flares coinciding in time with neutrino detections were noted before for two sources, PKS~1502+106 \citep[][has the highest activity index in our sample]{r:kielmann1502ATel} and TXS~0506+056 \citep[e.g.,][]{r:ros0506,r:kovalev0506}. Removing these two AGN in addition to the four strongest ones reduces the corresponding significance but the zero-lag peak in the correlation still remains pronounced at the highest frequency of 22~GHz.
This suggests that there are other sources behind the temporal correlation, even after dropping all those six.
We cannot reliably single out AGN responsible for this correlation as we did with the average flux density due to effectively less data but all of those with high values of $R_{\text{22GHz}}^{t=0}$ in \autoref{t:associations_ratan} are of interest for further more detailed studies.

\section{Interpretation and discussion} 
\label{s:discussion}

\subsection{Central Parsec-Scale Regions of Radio-Bright AGN as Probable Production Sites of High-Energy Neutrinos}
\label{s:discussion_core}

Complete samples of AGN selected on the basis of their parsec-scale flux density are dominated by jets observed at small viewing angles of a few degrees with typical Doppler factor in the range of 3 to 10 \citep[e.g.,][]{2019ApJ...874...43L}. Intrinsic opening angles of the jets are of the order of $1\degr$ \citep{JetAngles}.
Their VLBI flux density is dominated by the emission of the opaque core, which is the apparent base of the jet \citep[e.g.,][]{r:2cmfinescale,2012A&A...544A..34P}. The core is located at a typical deprojected distance on the order of 10~pc to the true nucleus \citep[e.g.,][]{2012A&A...545A.113P,2019MNRAS.485.1822P}. Its variability dominates that of the total radio flux density observed from AGN as confirmed observationally \citep[e.g.,][]{r:2cmfinescale} and follows from causality arguments.

Taken together with the results presented in the previous section, this implies that neutrinos are emitted in narrow beams pointing to the observer. As a result, it is possible to detect these neutrinos at the Earth from galaxies that reveal themselves as hosting bright Doppler-boosted parsec-scale jets.
We predict that analyzing VLBI-selected samples will allow researchers to find more of these AGN and associate neutrino production to their activity.
We note that similarly high VLBI flux density is crucial for $\gamma$-ray associations as shown by \citet{2009ApJ...707L..56K}.
The same beaming stands behind the $\gamma$-ray activity of some of VLBI-selected AGN \citep{2010A&A...512A..24S,2015ApJ...810L...9L,Stecker2019}, despite the fact that the origins of $\gamma$-rays and neutrinos may not be directly related, see \autoref{s:discussion:models}.
A possibility to detect AGN by VLBI, their $\gamma$-ray emission, or neutrinos requires a narrow viewing angle to the axis of the accretion disk~-- relativistic jet system. See also the discussion in \autoref{s:discussion:models}.

Significant correlation between the arrival dates of neutrino events and an increase of the total radio flux density is only seen at the highest frequency of 22~GHz in \autoref{f:xcorr}. This is easy to understand from the physics of AGN jet synchrotron radiation. Our temporal analysis deals with the flux density values around the neutrino event normalized by the average flux density outside this time range (\autoref{t:associations_ratan}).
The fraction of the core radiation and the relative strength of the core flares in total radio emission decrease with decreasing frequency \citep{1992ApJ...399...16A,2002PASA...19...83K,2016A&A...596A..45F}. This dependence is caused by the steep synchrotron spectrum of extended optically thin weakly variable jets and lobes \citep[e.g.,][]{2014AJ....147..143H}, as illustrated by analysis of continuum radio spectra \citep{1999A&AS..139..545K,2000PASJ...52.1027K,2002PASA...19...83K}.
Additionally, while the radio frequency decreases, the core flares peak with an increasing delay and have longer characteristic time scales due to the synchrotron opacity effect. 
Note that the peak at 11~GHz is weaker and slightly (about 1/4~yr) delayed relative to 22~GHz, as expected, although we do not assess statistical significance of this difference.
We expect that a similar temporal correlation with high-energy neutrino events could be delivered by an analysis of long-term single-dish monitoring observations of large enough VLBI-selected AGN samples by OVRO \citep{2011ApJS..194...29R} at 15~GHz and Mets\"ahovi Radio Observatory \citep{2004A&A...427..769T} above 20~GHz.

The observed temporal coincidence can be considered within the following scenario. An accreted material is accelerated to relativistic velocity close to the central super-massive black hole. As a result, high-energy protons are produced and generate neutrinos that begin their way to the observer. At the same time, a plasma blob starts propagating along the jet. It reaches the region where the jet is transparent to radio frequencies, and the observer starts seeing a delayed synchrotron flare. Observationally, we can limit the distance between the black hole and the transparent regions to be up to $\sim10$~pc from opacity arguments \citep{2012A&A...545A.113P,2019MNRAS.485.1822P}. Note that the typical apparent delay in the observer's frame is expected to be less than several months due to the small viewing angle of the jet. Compare with the discussion of the measured radio-$\gamma$-ray delay in AGN \citep[e.g.,][]{2010ApJ...722L...7P}.

The analysis in \autoref{s:an_avgflux} singles out four most probable sources of neutrino events in our sample. However, one can see from the values of $R$ in \autoref{t:associations_ratan} and from the comparison of two panels in \autoref{f:xcorr} that these sources did not have major month-scale flares at the time of the neutrino arrival. This might be explained by a strong emission of non-VLBI kpc-scale jet regions, which contribute to the total flux density measured by RATAN and smear out the radio activity index $R$. Another possible explanation is that the conditions in the cores of these strongest AGN are capable to produce observable neutrinos outside of the major flares. In the latter case, additional neutrino events from the directions of these sources, as well as from other sources with high VLBI flux densities might be expected. 3C\,279 is an interesting example with a low $R \approx 1$: we note that it underwent a non-major flare within several months from the IceCube event at 2015-09-26 \citep{2020MNRAS.492.3829L,2020arXiv200301999S}.

Once we have established that the central parsec-scale regions of radio-bright AGN are the production sites of at least a large part of the higher-energy ($E>200$~TeV) neutrinos detected by IceCube, we can now discuss implications of this observation for particular models of the neutrino origin.

\subsection{Implications for Models and the Lack of $\gamma$-ray Associations} \label{s:discussion:models}

The origin of high-energy astrophysical neutrinos from $p\gamma$ interactions in central parsecs of radio-bright active galaxies, supported by our study, has various theoretical grounds \citep[see, e.g.,][]{Sikora1990, Stecker:1991vm, Mannheim1, NeronovWhich, Stecker-1305.7404, Kalashev:2014vya}. At high energies, all interactions of energetic protons with ambient radiation or matter are dominated by the production of the lightest strongly interacting particles, $\pi$ mesons. They carry away most of the initial proton's energy. The probabilities to create one of three species of the mesons --- $\pi^0$,$\pi^+$, and $\pi^-$ --- are roughly equal. All mesons are unstable particles and decay: the energy of charged $\pi^\pm$ is carried out mostly by neutrinos while that of every $\pi^0$ is split between two photons. These physical processes are behind any non-exotic scenario of production of energetic astrophysical neutrinos above the proton rest energy of $\sim 1$~GeV. They are inevitably accompanied, at the production, by $\gamma$-rays of similar energies. Models of neutrino production in AGN, therefore, require either proton-proton ($pp$) or proton-photon ($p\gamma$) interactions \citep[see][for the earliest and the latest review, and further references]{EichlerSchramm1978,Cerruti-rev}. In the central regions of radio-loud AGN, $p\gamma$ interactions are always dominant because of strong radiation fields and relatively low target matter density \citep{Sikora1987}. The situation may be different in low-luminosity AGN or in large-scale jets but both are disfavored by our present results for energies above $\sim 200$~TeV. 

General features of the $p\gamma$ scenario \citep[see, e.g.,][and the references therein]{Boettcher-rev,Cerruti-rev} are derived from simple estimates for relevant particle-physics processes. At the energies we are interested in, the $p\gamma$ reaction goes dominantly through the resonant production of a $\Delta$ baryon. Consequently, $E_p'$ and $E_\gamma '$, the energies of $p$ and $\gamma$ in the frame of the production region, are related by $E_p' E_\gamma ' = m_\Delta^2$, where the $\Delta$ mass $m_\Delta \approx 1.232$~GeV. On the other hand, the kinematics of the $\pi$-meson production and decays provides the approximate relation for the energies of each of the three produced neutrinos in the same frame, $E_\nu ' \approx 0.05 E_p'$. These relations allow estimating the energies $E_p'$ and $E_\gamma '$ required to obtain a neutrino with the observed energy $E_\nu$ as follows:
$E_p'\approx 4\,\mbox{PeV}\,(E_\nu/200\,\mbox{TeV})\,(1+z)\,\delta^{-1}$,
$E_\gamma'\approx 411\,\mbox{eV}\,(200\,\mbox{TeV}/E_\nu)\,(1+z)^{-1}\,\delta$.
Here $z$ is the cosmological redshift, and $\delta$ is the Doppler factor of the region where neutrinos are produced.

The importance of the jet kinematics for the neutrino observations, cf.\ the $\delta$ factor in 
those equations,
is revealed by our study but was predicted long before the start of the IceCube observations. Ultrarelativistic momenta of accelerated protons are inherited by the reaction products, including neutrinos. In the observer's frame, they are additionally boosted by the Doppler factor of the jet bulk motion. This was pointed out by \citet{Mannheim1} and elaborated in detail, e.g., by \citet{NeronovWhich}. Interestingly, both papers composed short lists of potential neutrino-loud AGN. Each of them included one of the four brightest sources from our \autoref{t:associations}. 

The model of \citet{Mannheim1}, see also \citet{Mannheim2}, uses synchrotron target photons. 
This normally results in too high neutrino energies: to have enough energy to produce a $\pi$ meson on a soft synchrotron photon, the initial proton should also be too energetic itself. If this scenario explains the flux of neutrinos detected by IceCube at sub-PeV energies, then the flux at higher energies, above a few PeV, should be even higher. 
The predicted \citep{Mannheim3} diffuse flux from FSRQs is in a qualitative agreement with IceCube observations at 200~TeV. However, at ($10^4$--$10^5$)~TeV the predictions are too high to agree with IceCube \citep{IceCubeUHEnu} and Auger \citep{AugerUHEnu} upper limits on the diffuse neutrino flux. 

The synchrotron radiation from \emph{lower-energy} relativistic protons may contribute to the observed radio emission of compact jets in AGN \citep{Mannheim2,2000ARep...44..719K,r:kovalev0506} and be related to the high-energy protons that produce neutrinos. 
At the same time, explaining the full observed flux with this mechanism would require a total power of accelerated proton beam orders of magnitude higher than the Eddington luminosity of a supermassive black hole in the nucleus of an active galaxy \citep[see, e.g.,][]{Boettcher-p-blazar}. 
Contrary, models of \citet{Stecker:1991vm} and \citet{NeronovWhich}, further elaborated after first IceCube observations by \citet{Stecker-1305.7404} and \citet{Kalashev:2014vya}, use the emission from the accretion disk as the target for the $p\gamma$ interactions. This leads to the neutrino flux peaking at ($10^2$--$10^3$)~TeV, the energies of IceCube-detected neutrinos.

Next, the photon-photon cross section is two orders of magnitude larger than the $p\gamma$ one for relevant energies. This means that if $p\gamma$ interactions are efficient in a neutrino-production zone, then secondary energetic $\gamma$-rays, accompanying the neutrino production, interact even faster. They initiate electromagnetic cascades: energetic protons produce $e^+e^-$ pairs on the target photon background, then these electrons and positrons pass their energy again to ambient photons in inverse-Compton scattering. These upscattered photons produce pairs again, and the process continues until either the energy of photons falls below the pair-production threshold, or the interaction length becomes larger than the source size. These cascades, therefore, efficiently transfer the energy of the $\gamma$-ray photons down to the lower-energy band. No photons with energies similar to those of neutrinos can escape from the neutrino production region. Consequently, if strong $\gamma$-ray emission is observed from the same source, it may come from a different place than the neutrino emission and be produced by means of a different mechanism. Any connection between them is indirect \citep[see, e.g.,][and the reviews cited above]{b}. This is precisely what we observe in this study: the neutrino emission is found to be related to the parsec-scale radio flux density, whereas $\gamma$-ray fluxes of the radio-bright AGN from \autoref{t:associations} differ by orders of magnitude even for the four strongest radio sources that dominate the correlation. Explicit conclusions on the relation between $\gamma$-ray and neutrino fluxes depend crucially on the energies of target photons, 
cf.~equations above,
because they determine the pair-production threshold energy. These $E_\gamma'$ may differ significantly from one source to another. A quantitative study of this question is beyond the scope of the present work.

We see that our results agree well not only with the observational constraints but also with theoretical expectations.
In the scenario favored by our observations, there remains one important unconstrained element: how are the protons accelerated in the direct vicinity of a black hole? Among possible mechanisms are stochastic \citep{stochastic} or electrostatic \citep{magnetosphere-Rieger, magnetosphere-Neronov, magnetosphere-Istomin, magnetosphere-Ptitsyna} ones. Detailed quantitative modeling is necessary to understand whether the required proton energies can be obtained respectively in radio-bright AGN. Simple estimates show that energy losses to the neutrino-producing $p\gamma$ interactions limit the maximal energy of protons accelerated in the central regions of FSRQs by $\sim (10^4-10^5)$~TeV \citep{Sikora1987}. This is sufficient to explain even the highest-energy neutrinos detected by IceCube. These energies are, however, much lower than those required for ultra-high-energy cosmic rays (UHECR). This agrees well with the lack of the observed correlations between UHECR and neutrino arrival directions \citep[cf.][]{noUHECRcorr,noUHECRcorrWinter}. Kiloparsec-scale jets, lobes, and hot spots, as well as the central parts of low-luminosity active galaxies, are more probable UHECR acceleration sites \citep[see, e.g.,][]{PtitsynaUFN}.

\subsection{Relation to Previous Statistical Studies and Constraints} \label{s:discussion:previous}

A possible lack of direct relation between neutrino and $\gamma$-rays was formulated on general grounds in various previous studies and detailed for a particular blazar example TXS~0506+056 \citep[e.g.,][]{FedynitchNatureAstron,MuraseTXS}. These considerations help to understand the difference in the results between our study and a number of other correlation and stacking analyses aimed to figure out or to constrain plausible sources of IceCube neutrinos. Previous works mostly concentrated on $\gamma$-ray selected AGN as potential candidate neutrino sources \citep[e.g.,][]{corr-1601.06550Resconi, Kadler16, corr-1611.03874IceCube, corr-1611.06338Neronov, corr-1702.08779Vissani, corr-1807.04299Tavecchio, Krauss18, corr-1908.08458IceCube}. Constraints on the population of neutrino-emitting AGN \citep[e.g.,][]{MuraseCombined} also select them by their $\gamma$-ray luminosities. Contrary, our observations do not imply that
the $\gamma$-ray emission is necessarily a tracer of the neutrino emission and select potential sources by their VLBI radio flux.

Only in a few studies, arrival directions of high-energy astrophysical neutrinos were compared to the positions of AGN selected by other criteria than the $\gamma$-ray loudness. \citet{corr-Resconi-jets} used various criteria related to the estimated power of large-scale jets, testing, therefore, non-central parts of active galaxies, complementarily to our approach.
\citet{Krauss14,Kadler16} have studied VLBI properties and variability of AGN located within error regions of three PeV events with large positional uncertainties on the level of $10\degr$ (i.e., hundreds to thousand square degrees), which precludes highly significant associations to be made.
\citet{2017MNRAS.466L..34K} selected flat-spectrum radio quasars (FSRQ) by broadband radio-to-microwave spectral properties. Though the class of sources tested and the radio selection are common to their and to our study, our present work differs in the key point: the use of the VLBI flux density and a VLBI-selected statistically complete sample. None of the previous studies of IceCube neutrinos attempted to distinguish between central and extended parts of AGN, like we do here.

The previous stacking analyses discussed in this subsection are sensitive to the selection of $\gamma$-ray bright objects, and are, therefore, complementary to ours. There are statistical studies of another kind, which use neutrino events only. They are based on the search of small-scale anisotropy (clustering) of neutrino arrival directions, which allows one to constrain the number of sources contributing to the observed neutrino flux. See \citet{DubovskyClustering} for the description of the methodology in the cosmic-ray context, and \citet{MuraseCombined} for the most recent applications to high-energy neutrinos. Basically, if there are only a few neutrino sources on the sky, the strongest or nearest of them would reveal themselves by multiple events coming from the same direction. It is not the case for IceCube detections: the number of multiplets in the data is consistent with random fluctuations. This results in a lower bound on the number density of sources.
AGN as bright as those four selected in \autoref{s:an_avgflux} are relatively rare. There are only 26 AGN with a comparable level of historic VLBI flux denisty. This might seem to be in tension with clustering constrains. However, these constrains are relaxed if the most energetic $E>200$~TeV events are considered alone, like we do in this study. Moreover, radio quasars exhibit strong evolution with redshift, $\sim (1+z)^5$, which helps to relax the clustering constraints even further \citep{NeronovEvolution}.

We should mention one more general constraint on the population of high-energy neutrino sources. It is related to the accompanying photons co-created with neutrinos in $\pi$-meson decays. If not absorbed in the source, energetic photons initiate electromagnetic cascades, similar to those described in \autoref{s:discussion:models}, on the extragalactic background radiation \citep{Nikishov1962}. As a result, the energy initially emitted in the form of sub-PeV or PeV photons contributes to the diffuse gamma radiation in the $\sim 1-100$~GeV band, and is, therefore, constrained from above by the \textit{Fermi} LAT observations. These constraints disfavor transparent extragalactic objects as sources of $E \lesssim 100$~TeV neutrinos observed by IceCube. For a recent discussion, see, e.g., \citet{AhlersHalzen}. Independently on the assumption about the source opacity to $\gamma$-rays, these constraints are satisfied for the $E>200$~TeV hard component of the neutrino flux we study here.

We conclude that our results agree well with the previous observational studies and constraints.

\section{Summary} \label{s:summary}

The aim of the present study is to test whether high-energy neutrinos are produced in active galaxies and, if so, to localize neutrino-emitting regions within them. 
We analyze a set of 56 published IceCube events with directional errors less than 10~deg$^2$ and neutrino energies above 200~TeV.
It is found that AGN directionally coincident with neutrino events within statistical and systematic errors have, on average, higher historic VLBI flux density compared to other AGN within the all-sky complete flux-density-limited sample of 3388 sources.
We estimate the significance of this correlation by Monte-Carlo simulations and find the probability to observe the excess as a random fluctuation to be 0.2\,\%. This includes a correction for multiple trials related to the unknown value of the IceCube systematic error in arrival directions.
The four particular brightest sources that dominate the observed correlation are 3C\,279, NRAO~530, PKS~1741$-$038, and PKS~2145$+$067.

Further, we use the data from the RATAN-600 total radio flux density monitoring of VLBI-selected AGN and demonstrate that periods of increased emission at frequencies above 10~GHz correlate with neutrino detections. This result remains significant even when the four sources singled out by the average historic VLBI-flux-density analysis are removed from the sample. This means that other fainter AGN from the VLBI-selected sample are also neutrino emitters. In particular, the strongest flux density enhancement at the time of a neutrino event is observed for PKS~1502+106. This is a probable source of the 2019-07-30 IceCube event but is not among the four strongest objects discussed above.

For the first time, our study invokes the statistical power of radio observations to the problem of high-energy neutrino origin. 
We estimate systematic errors of IceCube directions and account for them in the analysis. The found systematic errors on the level of $0.5\degr$--$0.7\degr$ are compatible with the sparse published information.
VLBI turns out to be the key to the high-energy neutrino associations.
Altogether, these results suggest that a significant part of the observed $E \gtrsim 200$~TeV astrophysical neutrinos are produced in the central parsec-scale regions of radio-bright active galaxies with narrow Doppler-boosted relativistic jets pointing to the observer. 
These potential neutrino sources are found to have $\gamma$-ray fluxes that differ by more than two orders of magnitude. This is expected if the neutrino production region is opaque to energetic $\gamma$-rays due to pair-production cascades.

The results of our study support models in which protons are accelerated in collimated beams close to the central black hole of a powerful AGN and subsequently interact with ambient radiation from the accretion disk. Charged $\pi^\pm$ mesons born in these interactions pass their energy to $E \gtrsim 200$~TeV neutrinos eventually detected at the Earth, while accompanying neutral $\pi^0$ mesons decay to energetic photons that cascade down to lower energies in the same environment. 

These results may be used in quantitative modeling of the neutrino production taking into account their statistical nature. We explicitly list several AGN probably associated with neutrinos, and show that there are likely more in our sample. We estimate that there are only around 26 astrophysical neutrinos in the dataset considering the fraction of non-astrophysical background events (\autoref{s:data_icecube}). Thus, the AGN singled out by our analysis constitute a significant fraction of all high-energy astrophysical neutrino emitters, though other scenarios are not ruled out. In addition, the energy cut $E>200$~TeV selects only the highest-energy IceCube events, so we might reveal only one of several populations of neutrino emitters. 

Further observations will test and expand our findings. For our selection cuts of neutrino events, we expect about five IceCube alerts per year assuming the currently operated public alert system. Soon a similar number of track-like events will start coming from Baikal-GVD. Moreover, liquid-water experiments, Baikal-GVD and KM3NeT, will also provide cascade events with good angular resolution.
It is important to continue monitoring VLBI-selected AGN with single dish radio telescopes on a regular basis. Unfortunately, there is a global trend to finish such projects including the Michigan University program \citep{2017Galax...5...75A}, the F-GAMMA program \citep{2019A&A...626A..60A}, and possibly the OVRO program \citep{2011ApJS..194...29R}. Still ongoing programs include RATAN-600 \citep{2002PASA...19...83K}, Mets\"ahovi \citep{2004A&A...427..769T}, and POLAMI \citep{10.1093/mnras/stx2435}.
Dedicated VLBI monitoring observations of VLBI-compact AGN selected from within the neutrino positions immediately after alerts might help to directly relate neutrino production to parsec-scale properties of corresponding synchrotron flares.

\acknowledgments
We thank R.~Blandford, V.~Dokuchaev, A.~Fedynitch, A.~Franckowiak, S.~Gao, D.~Gorbunov, F.~Halzen, T.~Hovatta, M.~Kadler, D.~Levkov, E.~Lindfors, G.~Lipunova, K.~Murase, A.~Neronov, E.~Ros, V.~Rubakov, D.~Semikoz, C.~Spiering, F.~Stecker, O.~Suvorova, and the anonymous referee for helpful comments and discussions on various parts of this work.
We are grateful to E.~Bazanova for English language editing and proofreading of the text.
Observations at telescopes of the Special Astrophysical Observatory are supported by the Russian Ministry of Science and Higher Education (agreement 05.619.21.0016).
This study was supported in part by the Russian Science Foundation (project 16-12-10481).
This research has made use of NASA's Astrophysics Data System.
This research has made use of the NASA/IPAC Extragalactic Database (NED), which is operated by the Jet Propulsion Laboratory, California Institute of Technology, under contract with the National Aeronautics and Space Administration.
\facilities{IceCube neutrino observatory, VLBA, EVN, LBA, RATAN.}

\bibliography{neutradio}
\bibliographystyle{aasjournal}

\end{document}